\pdfoutput=1
\documentclass[aps,prl,twocolumn,floatfix,showpacs,superscriptaddress,footinbib]{revtex4}
\usepackage{amsmath,amssymb,bm,url,color,verbatim,psfrag}
\usepackage[ansinew]{inputenc}
\usepackage{subfigure}

\ifx\pdftexversion\undefined
  \usepackage[dvips]{graphicx}
\else
  \usepackage[pdftex]{graphicx}
\fi

\def \be{\begin{equation}}
\def \ee{\end{equation}}
\def \bew{\begin{widetext}\begin{equation}}
\def \eew{\end{equation}\end{widetext}}
\def \bmlett{\begin{mathletters}}
\def \emlett{\end{mathletters}}

\def \omegar{\omega_R}
\def \omegam{\omega_M}
\def \xrms{x_{\rm zpt}}

\def \nbar{\bar{n} }
\def \nosc{n_{\rm th}}
\def \nbath{n_{\rm bath}}

\def \SS{{\mathcal S}}

\def \ra{\rightarrow}

\def \hx{\hat{x}}

\def \hF{\hat{F}}
\def \hI{\hat{I}}

\def \hX{\hat{X}}

\def \hb{\hat{b}}
\def \hc{\hat{c}}
\def \hH{\hat{H}}
\def \hd{\hat{d}}

\def \hn{\hat{n}}

\def \bS{\bar{S}}

\def \Simp{\bS_{nn,{\rm imp} }}

\begin{document}

\title{Quantum Measurement of Phonon Shot Noise}

\author{A. A.\ Clerk}
\affiliation{Department of Physics, McGill University, Montreal, Quebec, Canada H3A 2T8}
\author{Florian Marquardt}
\affiliation{Department of Physics, Center for NanoScience, and Arnold Sommerfeld Center for Theoretical Physics, Ludwig-Maximilians-Universit\"at
M\"unchen\\ Theresienstr. 37, D-80333 M\"unchen, Germany}
\author{J. G. E.\ Harris}
\affiliation{Departments of Physics and Applied Physics, Yale University, New Haven, Connecticut 06520, USA}

\date{Feb 16, 2010}

\begin{abstract}
We provide a full quantum mechanical analysis of a weak energy measurement of a driven mechanical resonator.  We demonstrate that measurements
too weak to resolve individual mechanical Fock states can nonetheless be used to unambiguously detect the non-classical energy fluctuations of the driven mechanical resonator, i.e.~``phonon shot noise".  
We also show that the third moment of the oscillator's energy fluctuations provides a far more sensitive probe of quantum effects than the second moment, and that
measuring the third moment via the phase shift of light in an optomechanical setup directly yields the type of operator ordering postulated in the theory of full-counting statistics.
\end{abstract}

\pacs{
07.10.Cm, 
42.50.Lc. 
}

\maketitle

\textit{Introduction.}---  
There has been considerable interest recently in preparing and detecting quantum mechanical behaviour in mechanical resonators.  This general goal has been pursued using both optomechanical systems (where the mechanics is coupled to an optical cavity) and electromechanical systems (where the mechanics is coupled to an electrical circuit).  Not only do such studies attempt to test quantum mechanics in a new regime of large mass, they also have the potential of furthering our understanding of quantum dissipative processes and the boundary between classical and quantum physics.  Perhaps the simplest example of truly quantum behaviour is the energy quantization expected of a quantum oscillator.  Detecting this quantization directly by, e.g., observing quantum jumps between different Fock states \cite{Gabrielse99}, is extremely challenging \cite{Santamore04, Santamore04b, Harris08}.  One requires a detector which couples directly to the energy of the mechanical oscillator, not to its position $\hx$ as is more common; further, this coupling must be strong enough to resolve a mechanical energy quantum within the lifetime of a Fock state.  Recent optomechanical experiments \cite{Harris08} have demonstrated  coupling to $x^2$, which is equivalent to energy in the rotating wave approximation. However, it remains a challenge to satisfy all of the requirements for achieving a practical QND measurement of individual quantum jumps.  

As a result, it would be desirable to detect energy quantization 
in a mechanical oscillator using only the presently-existing resource of a detector that couples weakly to energy.  This is the goal of this Letter.  Similar to previous studies, we consider the QND measurement of the energy fluctuations of a dissipative mechanical resonator, but now consider the case where the mechanical resonator is strongly driven.  This drive will result in a large average number of quanta in the resonator, $\bar{n} \gg 1$.  
Our focus will be on the fluctuations of energy about this average value;  in the low-temperature limit, these fluctuations are completely quantum in nature and reflect the discreteness of the oscillator's energy.  
As the magnitude of this ``phonon shot noise"  scales with the magnitude of the applied drive, one can make it large enough to detect even if the detector-oscillator coupling is too weak to resolve individual Fock states.  We show that this is possible even in the presence of strong cavity cooling, which is necessary to ensure that the phonon shot noise dominates the thermal noise.  
We also analyze the fundamental backaction of the measurement, which manifests itself as mechanical frequency noise that can obscure the intrinsic quantum energy fluctuations.  We show that if the mechanical resonator is driven on resonance, the measurement is backaction evading, having a formal equivalence to a single-quadrature position measurement \cite{Braginsky80}.  As such, there is no fundamental quantum limit on the continuous monitoring of phonon shot noise.

Our study also reveals new physics associated with the energy fluctuations of a driven quantum resonator, namely that higher moments of the fluctuations are far more sensitive to the difference between the quantum and classical limit than the second moment.  The second moment in the zero-temperature quantum regime has a form that follows directly from the corresponding high-temperature, classical result:  one simply takes the classical expression and replaces the temperature $T$ by $\hbar \omegam / 2 k_B$, where $\omegam$ is the mechanical mode frequency.  
We show that this simple correspondence {\it does not} hold for higher moments of the energy fluctuations.

\textit{Model.}---
We consider a generic setup in which a detector is weakly coupled to the energy of a damped, driven mechanical resonator.  We let $\hF$ denote the detector quantity which directly couples to the oscillator number operator $\hn = \hc^\dag \hc$, and $\hI$ denotes the detector quantity that is directly monitored. We take $\hbar = 1$.  The Hamiltonian has the form:
\begin{eqnarray}
	\hH =  \omegam \hc^\dag \hc  + \hH_{\gamma}  -  f \left( e^{i \omega_D  t} \hc + h.c. \right) + \hH_{\rm det} + \hF \cdot \hn
	\label{eq:Hamiltonian}
\end{eqnarray}
Here $\hH_{\gamma}$ describes the damping (at a rate $\gamma$) of the mechanical resonator by a thermal oscillator bath, $f$ is the magnitude of the coherent oscillator driving force (frequency $\omega_D = \omegam + \delta$), and $\hH_{\rm det}$ is the detector Hamiltonian.  In the absence of thermal noise, the coherent oscillator drive would
yield an average number of mechanical quanta $\langle \hn \rangle \equiv \nbar = 4 f^2 / ( 4 \delta^2 + \gamma^2 )$. 
A concrete example of such a system is the ``membrane-in-the-middle" optomechanical system discussed in Refs.~\cite{Harris08, Jayich08, SankeyDraft2010}.  The detector here is a driven optical cavity whose frequency $\omegar$ depends quadratically on the displacement of a dielectric membrane placed in the cavity; this membrane is the mechanical oscillator.  Within a rotating-wave approximation, one obtains the desired coupling to the oscillator's energy, with the input operator given by $\hF = A \hn_{\rm cav}$, where $\hn_{\rm cav}$ is the cavity photon number and $A = (d^2 \omegar / dx^2) \xrms^2$ is the quadratic optomechanical coupling.  The output operator $\hI$ here would correspond to the error signal in a Pound-Drever-Hall scheme used to monitor the cavity frequency (and hence 
$\hn$).

For the weak coupling situation we focus on, linear response theory applies, implying $\langle \hI (t) \rangle = \lambda \langle \hn(t) \rangle$, where $\lambda$ is the detector response coefficient.
We focus on the experimentally-relevant limit where the oscillator energy evolves slowly enough that we can ignore the frequency dependence of $\lambda$.  A straightforward calculation yields that the symmetrized noise spectrum of $\hI$ at low frequencies is given by:
\begin{subequations}
\begin{eqnarray}
\bS_{II}[\omega] & = & \lambda^2 \left( \bS_{nn,\rm imp} +  \bS_{nn}[\omega]  +  \bS_{nn,{\rm BA}}[\omega] \right)
	\label{eq:SII}\\
\bS_{nn}[\omega] & = &
	\bar{n} \left(1 + 2 \nosc \right) \left( \rho[\omega+\delta] + \rho[\omega-\delta] \right) 
	\label{eq:Snn}
\end{eqnarray}
\end{subequations}
where $\rho[\omega] = (\gamma/2) / (\omega^2 + \gamma^2/4)$, and $\gamma$ is the total damping of the mechanical oscillator.  We have anticipated using a large magnitude oscillator driving force, and have thus only retained oscillator terms proportional to $\nbar$.
The first term in Eq.~(\ref{eq:SII}), $\Simp$, describes the intrinsic output noise floor of the detector (i.e.~the imprecision noise), while the second term corresponds to the amplified number fluctuations $\bS_{nn}$ of the mechanical resonator.  These number fluctuations (c.f.~Eq.~(\ref{eq:Snn})) have both a quantum shot-noise part which is non-vanishing at zero-temperature and a classical part proportional to the effective number of thermal quanta in the oscillator, $\nosc$.  This thermal contribution corresponds to the amplification of the thermal force noise driving the oscillator by the coherent drive.
Finally,  the third term in Eq.~(\ref{eq:SII}) describes a backaction contribution to the output spectrum:  by virtue of the detector-oscillator coupling in Eq.~(\ref{eq:Hamiltonian}), fluctuations of $\hF$ will result in fluctuations of the mechanical oscillator frequency, thus enhancing its number fluctuations and yielding extra noise in the output.

\textit{Resolving phonon shot noise.}---
We first discuss the resolvability of the phonon shot noise ignoring the effects of backaction.  A first requirement is to have the quantum shot noise contribution to $\bS_{nn}$ overwhelm the classical, thermal contribution.  We thus require a cold oscillator,  $\nosc \ll 1$.  In the optomechanical setup, this could be achieved by using a second optical mode whose linewidth is smaller than $\omegam$ to laser-cool the mechanical mode, as discussed in Refs.~\cite{Braginsky02, Marquardt07, WilsonRae07};
such a simultaneous use of different optical modes for cooling and measurement can be achieved in the device of Ref.~\cite{SankeyDraft2010}.
The use of laser-cooling comes at a price:  it increases the total mechanical oscillator damping compared to its intrinsic value $\gamma_0$.  This in turn reduces the resolvability of the oscillator peak in the output spectrum.  One has the simple relation $\gamma = \gamma_0 (\nbath / \nosc )$, where $\nbath$ is the bath temperature (expressed as a number of quanta).

Resolving the peak associated with $\bS_{nn}[\omega]$ in the output spectrum corresponds to continuously measuring the phonon shot noise. 
Even without backaction effects, it is not clear that this can be accomplished, given both that strong cooling is required, and that the intrinsic detector-oscillator coupling is weak. 
Consider the case $\delta = 0$, which gives a maximal oscillator-induced peak in $\bS_{II}[\omega]$.  As a measure of the resolvability of the phonon shot noise, we consider the peak-to-noise ratio $\mathcal{S} \equiv \bS_{nn}[0] / \Simp$.  In the interesting limit
$\nosc \ra 0$, $\bar{n} \gg 1$,  we find:
\begin{eqnarray}
	\mathcal{S} = \frac{4   \nbar / \gamma}{ \Simp } = \frac{4   \nbar}{ \Simp  \gamma_0 } \times \frac{\gamma_0}{\gamma} = 8 \bar{n} \nosc \Sigma^{(0)}
	\label{eq:Sratio}
\end{eqnarray}
Here, $\Sigma^{(0)} = 1/(2 \gamma_0 n_{\rm bath} \Simp)$ is the signal-to-noise ratio introduced in Ref.~\cite{Harris08} associated with resolving a quantum jump of the mechanical resonator from its ground to first excited state.  The fact that $\mathcal{S} \propto \nosc$ reflects the increase in $\gamma$ associated with cavity cooling.  We see that with a suitably strong mechanical drive, $\mathcal{S} > 1$ even though the measurement strength is weak (i.e.~$\Sigma^{(0)} < 1$) and though a significant amount of backaction cooling is required (i.e.~$\nosc \ll 1 \ll \nbath$).

We can easily apply our general analysis
to the membrane-in-the-middle device demonstrated in Ref.~\cite{SankeyDraft2010}. This device is similar to those described in Refs.~\cite{Harris08,Jayich08}, but achieves values of $\partial^2 \omega_R / \partial x^2$ up to three orders of magnitude greater 
while maintaining negligible optical absorption\cite{SankeyDraft2010}. 
The parameters of this device are listed in Table 1. 
We assume the device is pre-cooled inside a cryostat to a bath temperature $T = 300$ mK; this would ensure that laser cooling to $n_{\mathrm{th}} < 1$ is readily feasible \cite{Marquardt07, WilsonRae07}. We also assume the membrane is driven to an amplitude of 2 nm. This is below the onset of dynamical bistability in these devices \cite{Zwickl2008}, and corresponds to mean phonon number $\bar{n} = 4.76 \times 10^{12}$.  The resulting peak-to-noise ratio is $\SS > 20$.
Thus, by combining the measurement scheme presented here with the enhanced $\partial^2 \omega_R / \partial x^2$ demonstrated in Ref.~\cite{SankeyDraft2010}, it should be possible to observe energy quantization in an existing device. This is in sharp contrast to the proposal for detecting individual phonons in an optomechanical device, which would require substantial improvements to the membrane, the cavity, and their coupling.\cite{Harris08}

\begin{table*}
\caption{Optomechanical device parameters. $m, \omega_m, Q$: membrane mass, resonance frequency, and quality factor. $F, L, P_{\mathrm{in}}$: cavity finesse, length, and input power. $T$: bath temperature. With the exception of $T$ and $n_{\mathrm{th}}$ these parameters are those of the device demonstrated in Ref.~\cite{SankeyDraft2010,  Zwickl2008}.  They allow a peak-to-noise ratio $\SS=20.6$ for the phonon shot noise measurement.}
\begin{ruledtabular}
\begin{tabular}{llllllllll}
	$m$: 40 ng	&
	$\frac{\omega_m}{2 \pi}$: $1$ MHz	&
	$Q$: $1.2 \times 10^7$	&
	$F$: $5 \times 10^4$	&
	$L$: $67$ mm	&
	$P_{\mathrm{in}}$: 10 $\mu$W	&
	$T$: 300 mK	&
	$\frac{\partial^2 \omega_R}{ \partial x^2}$: 3 MHz/$\rm{nm}^2$	&
	$n_{\mathrm{th}}$: 0.2	
\end{tabular}
\end{ruledtabular}
\end{table*}


\

\textit{Backaction.}---
We now consider the effects of the fundamental measurement backaction.  These result from fluctuations of the input operator $\hF$, which lead to frequency noise of the mechanical oscillator and enhanced mechanical energy fluctuations. 
One might expect that to prevent this backaction contribution from obscuring the phonon shot noise signature, there will be a limit to how large one can make $\nbar$, and hence a limit on the maximum signal to noise ratio $\mathcal{S}$ in Eq.~(\ref{eq:Sratio}).  A similar situation arises in the continuous monitoring of zero-point position fluctuations, where the analogous peak-to-noise ratio $\mathcal{S} \leq 3$ \cite{Clerk08b}.

To analyze the effects of backaction in the large $\nbar$ limit of interest, we let $\hd$ denote the fluctuations of the mechanical oscillator annihilation operator from its average value:  $\hc(t) = e^{-i \omega_D t} \left(  \sqrt{\bar{n}} e^{i \beta} + \hat{d}(t) \right)$.  In the rotating frame, the mechanical oscillator Hamiltonian takes the form $\hH_{\rm M} =  -\delta \hd^{\dag} \hd$. 
Further, to leading order in $\nbar$, the oscillator number operator becomes $\hat{n}(t) \simeq \nbar + \sqrt{\nbar} \hat{X}(t)$
and the oscillator-detector coupling Hamiltonian $\hH_{\rm int} = \sqrt{\nbar} \hat{F} \hat{X}$, where $\hat{X}(t) = \left(e^{-i \beta} \hd(t) +  
e^{i \beta} \hd^{\dag}(t) \right)$.  Measuring the oscillator energy thus corresponds to a measurement of the ``position" operator $\hat{X}$, and a continuous measurement of the phonon shot noise to a measurement of the zero-point fluctuations of $\hat{X}$.  As with standard continuous position detection, there will in general be a backaction associated with this measurement, as $\hX$ does not commute with $\hH_{\rm M}$.  However, in the special case of a resonant oscillator drive (i.e.~$\delta = 0$), $\hH_{\rm M}$ vanishes.  $\hat{X}$ is thus trivially a QND observable that can be measured without a backaction limit.  Thus, for an on-resonant drive, and to leading order in $\nbar$, there is no backaction associated with the phonon shot noise measurement; we thus anticipate that there is no backaction-induced quantum limit on $\mathcal{S}$.    Formally, the measurement of phonon shot noise in this large-$\nbar$ limit corresponds to a ``QND in time" measurement of a single motional quadrature \cite{Braginsky80}.  

To verify that there is indeed no backaction limit, we need to describe the leading non-vanishing backaction contribution when $\delta = 0$.   This arises from the term $\hd^\dag \hd$ in $\hn$ that was neglected above.  Treating the fluctuations of $\hF$ as being Gaussian (as is appropriate for the weak couplings we consider),  we find that the leading-order backaction-driven number fluctuations are given by $\bS_{nn,{\rm BA}}[0] = 	 \frac{4 \nbar^2}{\gamma} \left( \frac{ \bS_{FF} }{ \gamma } \right)^2$.  The total added noise of the measurement includes this contribution, plus the imprecision noise of the detector (first term in Eq.~(\ref{eq:SII})).  The total added noise can be represented as an effective number of thermal oscillator quanta $n_{\rm add}$ via $n_{\rm add} = \left( \Simp + S_{nn,{\rm BA}}[0] \right) / (2 S_{nn}[0,\nosc=0] )$.  
 Assuming our detector has quantum-limited noise properties and thus optimizes the Heisenberg noise inequality $\Simp \bS_{FF} \geq 1/4$ \cite{Clerk08b}, we find that for a fixed value of $\nbar$, the minimal possible value of $n_{\rm add}$ is given by:
\begin{eqnarray}
	n_{\rm add} \Big|_{\rm min} =  \frac{3}{16 \left( 2\nbar \right) ^{1/3} } 
\end{eqnarray}
where the minimum is achieved for an optimal measurement strength satisfying  $\Simp \cdot \gamma = (32 \nbar^2)^{1/3}$.  We thus see that even with the inclusion of backaction effects, the added noise of the measurement (referred back to the oscillator) can be made arbitrarily small by driving the oscillator on-resonance with a sufficient strength.
Note that this is not true if one drives the oscillator off-resonance.  In this case, even in the limit $\nbar \ra \infty$, the added noise $n_{\rm add}$ cannot be made smaller than $| \delta / \gamma|$.  The device described in Table 1 is not optimally coupled, but backaction effects nonetheless only yield $n_{\rm add} \simeq  4 \times 10^{-5}$.

Finally, in the optomechanical realization of our scheme, one must also consider the backaction mechanism analyzed in Ref.~\cite{Miao09}.  This mechanism is absent in the ideal case of a one-port cavity; in the more general case of a two-port cavity, the constraint set by this mechanism on the cavity damping is much weaker for our scheme than for a single-phonon measurement, as our scheme utilizes much smaller optomechanical coupling.

\textit{Non-classical higher moments.}---
 While the phonon shot noise described by the zero-temperature limit of $S_{nn}[\omega]$ is a completely quantum phenomenon, its form is not so different from the classical, high-temperature 
prediction for the energy fluctuations of a driven oscillator.  The quantum, zero-temperature limit of $S_{nn}[\omega]$  is simply obtained by taking the corresponding high-temperature classical expression (second term in Eq.~(\ref{eq:Snn})), and replacing $\nosc$ with $0.5$.  In other words, the quantum shot noise is identical to the classical expression evaluated at a temperature $T = \hbar \omega / 2 k_B$.   The energy fluctuations of a driven oscillator are however not Gaussian in either the classical or quantum limit; thus, a natural question is whether such a simple correspondence between the two limits also exists for the higher moments.  We now show that this is not the case by calculating the third moment of the driven oscillator's energy fluctuations.  

For simplicity, we consider the fluctuations of the time-integrated phonon number $\hat{m} = \int_0^t dt' \hn(t')$ in the long-time limit.  Experimentally, one would attempt to measure this quantity by time-integrating the detector output $I(t)$; its mean is simply given by $\langle \hat{m} \rangle = \langle \hn \rangle t$.  In considering the second and higher moments of $\hat{m}$ quantum mechanically, care must be taken to account for the fact that $\hn(t)$ operators at different times do not commute.  One must treat the measurement quantum mechanically (not just the oscillator) to extract the proper definition of the higher-moments as measured in the experiment.  

The above ordering problem has been addressed extensively in the study of the counting statistics of electron transport through phase coherent conductors.  Several different models of an ideal measurement yield the same definition for the higher moments, the so-called Keldysh ordering \cite{Levitov96, Nazarov03b}.  For the second central moment, one obtains in the long-time limit $\langle \langle \hat{m}^2 \rangle \rangle = \bS_{nn}[0] t$, as would be obtained directly from the definition of $\hat{m}$.  However, for the third central moment, a far less obvious answer is given ($\delta \hn \equiv \hn - \langle \hn \rangle$):
\begin{eqnarray}
	\langle \langle \hat{m}^3 \rangle \rangle_K &= &   
		\int_0^{t} dt_1 \int_0^{t} dt_2 \int_0^{t} dt_3
		g 
			\langle \delta \hn(t_1) \delta \hn(t_2) \delta \hn(t_3) \rangle
		 \nonumber \\
	g  & = &   \frac{3}{2} \left[1 - \theta(t_1 - t_2) \theta (t_3 - t_2) \right] 
	\label{eq:ThirdMomentDefinition}
\end{eqnarray}
The classical definition would simply have $g = 1$; instead, the quantum answer involves neglecting contributions to the integral where the middle $\hn$ operator appears at the earliest time.  This ordering follows from a consideration of how the measured operator $\hn$ influences the evolution of the density matrix of the detector  \cite{Levitov96, Nazarov03b}.  Importantly, one can derive Eq.~(\ref{eq:ThirdMomentDefinition}) explicitly for the concrete optomechanical realization of our system, where the mechanical mode energy modulates the frequency of a resonantly-driven single-port optical cavity.  To see this, we assume that the output field from the cavity is subjected to homodyne detection, and that the mechanical damping is much smaller than the cavity damping.  Using standard input-output theory \cite{Clerk08b}, the output field from the mixer used in the homodyne setup will have the form $\hb(t) = \beta + B \hn(t)$, where $\beta$ parameterizes the large magnitude of the classical reference beam used, $B$ is a constant proportional to the optomechanical coupling,
and we omit terms describing shot noise. 
The intensity $\hI = \hb^\dag \hb$ is then measured using a photodetector; $\delta \hat{I} =  \hb^\dag \hb - \beta^2$ is the output of the measurement. 

Experimentally, one would try to extract the third moment of $ \int dt \delta \hI(t)$ from
the observed third moment of the number of photons detected in the output port of the homodyne setup within a given time-interval. To calculate this, one has to use Glauber photodetection theory\cite{Glauber63}. In particular, the third-order intensity correlation will be given by
$	\langle :I(t_1) I(t_2) I(t_3):\rangle  = 
		\langle \hb^\dag(t_a) \hb^\dag(t_b) \hb^\dag(t_c) \hb(t_c) \hb(t_b) \hb(t_a) \rangle$,
where $t_a < t_b < t_c$ denotes the time-ordered listing of $t_1,t_2$ and $t_3$.  
By thus calculating the leading contribution (in $\beta$) to the third moment of the number of detected photons in the output port of the homodyne setup, one directly recovers the Keldysh ordering described by Eq.~(\ref{eq:ThirdMomentDefinition}).
To our knowledge, this is the first demonstration of how the Keldysh ordering arises in a realistic measurement setup.

Applying this definition to our system, we find:
\begin{eqnarray}
	\langle \langle m^3 \rangle \rangle = 
		\frac{6 \nbar t}{\gamma^2} \left[ 
		\frac{ 3 - 4 (\delta/ \gamma)^2 + 16 \nosc (\nosc+1) }{ \left(1 + 4 \delta^2 / \gamma^2 \right)^2} \right]
		\label{eq:OscillatorThirdMoment}
\end{eqnarray}	
We have neglected a purely thermal contribution to Eq.~(\ref{eq:OscillatorThirdMoment}) which is independent of $\nbar$.
In the high-temperature classical limit (i.e.~$\nosc \gg 1$), the skewness of $\hat{m}$ is always positive, and the dependence on the detuning $\delta$ of the mechanical drive enters only through the oscillator's susceptibility.  In contrast, in the quantum limit $\nosc \ra 0$, the skewness can be positive or negative depending on $\delta$.  Thus, the third moment of energy fluctuations has a much greater sensitivity to whether one is in the quantum or classical limit:  the quantum expression does not simply correspond to evaluating the classical expression at an effective temperature $T = \hbar \omega_M / 2 k_B$.  Note that the use of the proper Keldysh ordering is crucial to obtaining Eq.~(\ref{eq:OscillatorThirdMoment}).  If one simply took $g=1$ in Eq.~(\ref{eq:ThirdMomentDefinition}), one would get the correct high-temperature limit, but would erroneously find that the zero-temperature expression corresponds to the classical expression evaluated at  $T = \hbar \omega_M / 2 k_B$.  Note that the full distribution of the driven oscillator's energy fluctuations can be directly obtained using the approach of Ref.~\cite{Clerk07}.

\begin{figure}
\begin{center}
\includegraphics[width = 0.99 \columnwidth, angle=0]{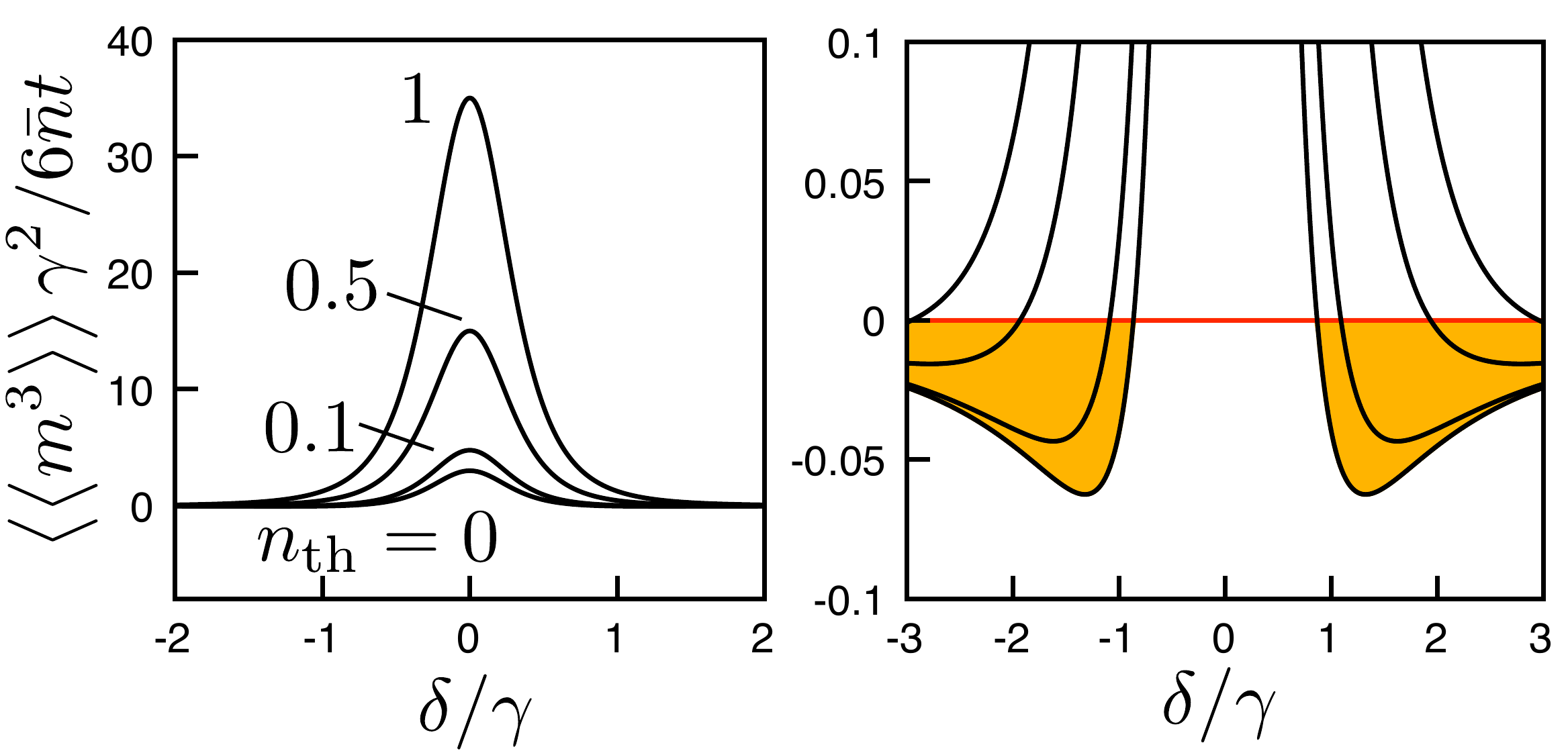}
\vspace{-0.60 cm}
\caption{(a) Third moment $\langle \langle m^3 \rangle \rangle$ of integrated energy fluctuations resulting from phonon shot noise, as a function of the drive detuning $\delta$.  (b)  Zoom of plot in (a), showing that at low temperatures, $\langle \langle m^3 \rangle \rangle$ can become negative in the quantum limit.}
\label{fig:Fig1}
\end{center}
\vspace{-0.90 cm}
\end{figure}
%

We thank J.C. Sankey for discussions.  A.C. acknowledges funding from CIFAR, NSERC and the Alfred Sloan Foundation; F.M. from the DFG (Emmy-Noether, SFB 631, NIM), GIF and DIP; J.G.E.H. from NSF and AFOSR. This material is based upon work supported by DARPA under Award No. N66001-09-1-2100

\bibliographystyle{apsrev}
\bibliography{ACTotalRefs}

\end{document}